# Appearance of Successive Phase Transition in SmRu$_4$P$_{12}$ under High Magnetic Fields Probed by $^{31}$P Nuclear Magnetic Resonance

Kenichi HACHITANI$^{1,2}$*, Hideaki AMANUMA$^1$, Hideto FUKAZAWA$^{1,3}$, Yoh KOHORI$^{1,3}$, Keiichi KOYAMA$^4$, Ken-ichi KUMAGAI$^5$, Chihiro SEKINE$^6$ and Ichimin SHIROTANI$^6$

$^1$*Graduate School of Science and Technology, Chiba University, Chiba 263-8522*
$^2$*Advanced Meson Science Laboratory, RIKEN (The Institute of Physical and Chemical Research), Wako, Saitama 351-0198*
$^3$*Department of Physics, Faculty of Science, Chiba University, Chiba 263-8522*
$^4$*High Field Laboratory for Superconducting Materials, Institute for Materials Research, Tohoku University, Sendai 980-8577*
$^5$*Division of Physics, Graduate School of Science, Hokkaido University, Sapporo 060-0810*
$^6$*Department of Electrical and Electronic Engineering, Muroran Institute of Technology, Muroran, Hokkaido 050-8585*

The $^{31}$P-NMR (nuclear magnetic resonance) measurements of the filled skutterudite SmRu$_4$P$_{12}$ have been carried out in several applied magnetic fields. The line width of the $^{31}$P-NMR spectrum rapidly increases below the metal-insulator transition temperature $T_{\rm MI}$, which indicates the appearance of an internal field below the temperature. Though no distinct anomaly was observed below $T_{\rm MI}$ in low fields, a complicated structure was observed between the Néel temperature $T_{\rm N}$ and $T_{\rm MI}$ ($T_{\rm N} < T_{\rm MI}$) above 70 kOe. The spin-lattice relaxation rate $1/T_1$ is almost independent of temperature above $T_{\rm MI}$, and rapidly decreases below $T_{\rm MI}$. The $1/T_1$ in low fields decreases monotonously, whereas the strong suppression of $1/T_1$ occurs between $T_{\rm N}$ and $T_{\rm MI}$ above 70 kOe. The structure of the spectra and the suppression of $1/T_1$ become apparent with increasing field.

## 1. Introduction

Filled skutterudites $RT_4X_{12}$ ($R$: rare earth, actinide; $T$: Fe, Ru, Os; $X$: P, As, Sb) show a wide variety of physical properties.[1–7] Remarkable phenomena observed in these systems are caused by the combination of the strong hybridization effect between the p-conduction and the f-localized electrons, the unique band structure, the orbital degree of freedom of the $R$ ions and the crystalline electric field (CEF) effect. Among them, PrRu$_4$P$_{12}$ and SmRu$_4$P$_{12}$ exhibit metal-insulator (M-I) transitions at the temperatures $T_{\rm MI}$ of 62 K and 16.5 K, respectively.[5,8] The M-I transition in PrRu$_4$P$_{12}$ is a nonmagnetic one caused by the band nesting. On the other hand, the M-I transition in SmRu$_4$P$_{12}$ accompanies a magnetic anomaly. Different features of each M-I transition have attracted much attention.

The $T_{\rm MI}$ of SmRu$_4$P$_{12}$ exhibits the characteristic magnetic field dependence.[9,10] It increases with increasing field up to 200 kOe, and saturates around 300 kOe. The reentrant behavior is expected at higher fields. In addition, it is considered that an extra phase transition occurs at 15 K in zero field.[9,10] The transition temperature which is called $T_{\rm N}$ decreases monotonously with increasing field. Though the associate anomaly is indistinct in low fields, it gradually becomes apparent with increasing field. The behavior is similar to that of CeB$_6$ where an antiferro-quadrupolar (AFQ) order and a subsequent antiferromagnetic (AFM) order occur.[11] Hence, the successive transition in SmRu$_4$P$_{12}$ had been expected to be the AFQ order below $T_{\rm MI}$ (the phase I above $T_{\rm MI}$ and the phase II between $T_{\rm N}$ and $T_{\rm MI}$) and the AFM order below $T_{\rm N}$ (phase III), respectively.

On the contrary, recent ultrasonic measurements have shown that the elastic constant of SmRu$_4$P$_{12}$ exhibits a slight and a large softening above $T_{\rm MI}$ and below $T_{\rm MI}$ toward $T_{\rm N}$, respectively.[12,13] The softening above $T_{\rm MI}$ is much smaller than that expected for the case of the AFQ order where the quadrupole-quadrupole interaction plays an important role.[13] From group theoretical considerations, the octupolar ($\Gamma_{5u}$) order below $T_{\rm MI}$ and the AFM ($\Gamma_{4u}$) one below $T_{\rm N}$ are the most probable candidates for the order parameters of the phase II and III, respectively.[13] There is a clear difference between the AFQ order and the octupolar order in zero field. The AFQ order is non-magnetic and holds the time reversal symmetry. On the contrary, the octupolar order is magnetic one which spontaneously breaks the symmetry. Zero field and longitudinal fields $\mu$SR measurements have shown the appearance of a static internal field below $T_{\rm MI}$, which excludes the occurrence of the AFQ order in the phase II.[14,15] Indeed, it is theoretically pointed out with multi-orbital Anderson model calculations that the $\Gamma_{5u}$ and the $\Gamma_{4u}$ octupolar fluctuations can become significant in Sm-based filled skutterudite systems.[16]

It is required to determine the order parameters of the phase II and III in SmRu$_4$P$_{12}$. In order to clarify the physical properties of this system, it is important to understand the CEF effect. Here, we have carried out $^{31}$P-NMR (nuclear magnetic resonance) of SmRu$_4$P$_{12}$ in several applied magnetic fields in order to investigate the system microscopically. In this paper, we report the results of the measurements below $T_{\rm MI}$ including that at high temperatures above $T_{\rm MI}$.

*E-mail address: hachitani@physics.s.chiba-u.ac.jp

## 2. Experimental

The single-phase polycrystalline $SmRu_4P_{12}$ was synthesized by using the high temperature and high pressure method.[8] The sample was crushed into powder for the experiments. The $^{31}$P-NMR experiments have been carried out by using phase-coherent pulsed NMR spectrometers and superconducting magnets. The NMR spectra were measured both by sweeping the applied fields at a constant resonance frequency and by sweeping the resonance frequency at a constant applied field. The spin-lattice relaxation rate $1/T_1$ was measured with the saturation recovery method in applied fields. The nuclear magnetization recovery curve was fitted by a simple exponential function as expected for the nuclear spin $I = 1/2$ of the $^{31}$P nucleus. The experiment in the field of 150 kOe was performed at High Field Laboratory for Superconducting Materials, Institute for Materials Research, Tohoku University.

## 3. Results

### 3.1 $^{31}$P-NMR spectra above $T_{MI}$

Figure 1 shows the frequency swept $^{31}$P-NMR spectra at several temperatures $T$ above $T_{MI}$ in the field of 94.063 kOe. The line width of the spectra increases with decreasing temperature and with increasing field. The spectra above about 70 K show a typical powder pattern with the uniaxial Knight shift distribution for the nuclear spin $I = 1/2$. Below about 70 K, the uniaxial symmetry gradually decreases with decreasing temperature. The $^{31}$P-NMR spectra were analyzed by taking into account of the powder average of the Knight shift $K$ anisotropy and the excess Gaussian broadening. The $K$ is expressed as

$$K = K_{iso} + K_{ax}(3\cos^2\theta - 1) - K_{aniso}(\sin^2\theta \cos 2\phi), \quad (1)$$

where $K_{iso}$, $K_{ax}$ and $K_{aniso}$ represent an isotropic term, a uniaxial term and the deviation from the uniaxial symmetry, respectively.

Figure 2 shows the bulk susceptibility $\chi$ dependence of $K_{iso}$ ($K$-$\chi$ plot). The $K_{iso}$ almost follows the liner function of the $\chi$ expressed by the formula

$$K_{iso} = \frac{A_{hf}}{N_A \mu_B}\chi, \quad (2)$$

where $N_A$ and $\mu_B$ are the Avogadro constant and the Bohr magneton, respectively. The hyperfine coupling constant $A_{hf} \simeq 2.7$ kOe/$\mu_B$ was estimated from the slope of the solid line shown in Fig. 2. This value is consistent with a previous result.[17]

Figure 3 shows the $T$ dependence of $K_{ax}$ and $K_{aniso}$. The $K_{ax}$ has the weak $T$ dependence in the whole $T$ region with the value of about 0.013 %. The $K_{aniso}$ decreases with increasing temperature below about 70 K, and decreases very gradually at higher temperatures (0.0044 % at 292 K). One origin of the anisotropic hyperfine coupling is the the direct dipoler coupling of $Sm^{3+}$ magnetic moments and $^{31}$P nuclear spins. The lattice sum calculation predicts nearly uniaxial Knight shift to be $K_{ax} \simeq 0.007$ % and $K_{aniso} \simeq 0.0008$ % at 292 K. These results are shown in Table I including the experimental results. This indicates that half of the anisotropic hyperfine coupling can be explained by the direct dipolar coupling of the $Sm^{3+}$ magnetic moments and the $^{31}$P nuclear spins. An anisotropic field via the dipolar coupling of P-3p spins and $^{31}$P nuclei induced by spin polarizations at the P-3p spins makes an extra contribution for the anisotropic hyperfine coupling.[18] The sum of these contributions would explain the observed values. The change of the $^{31}$P-NMR spectra below about 70 K is represented by the increase of $K_{aniso}$, which would be due to that the change of Sm-4f orbital shape occurs associated with the thermal excitation of the CEF levels. This is consistent with the result that the CEF level splitting between the ground state $\Gamma_5$ and the excited $\Gamma_{67}$ of 60 K was estimated from a specific heat measurement.[19]

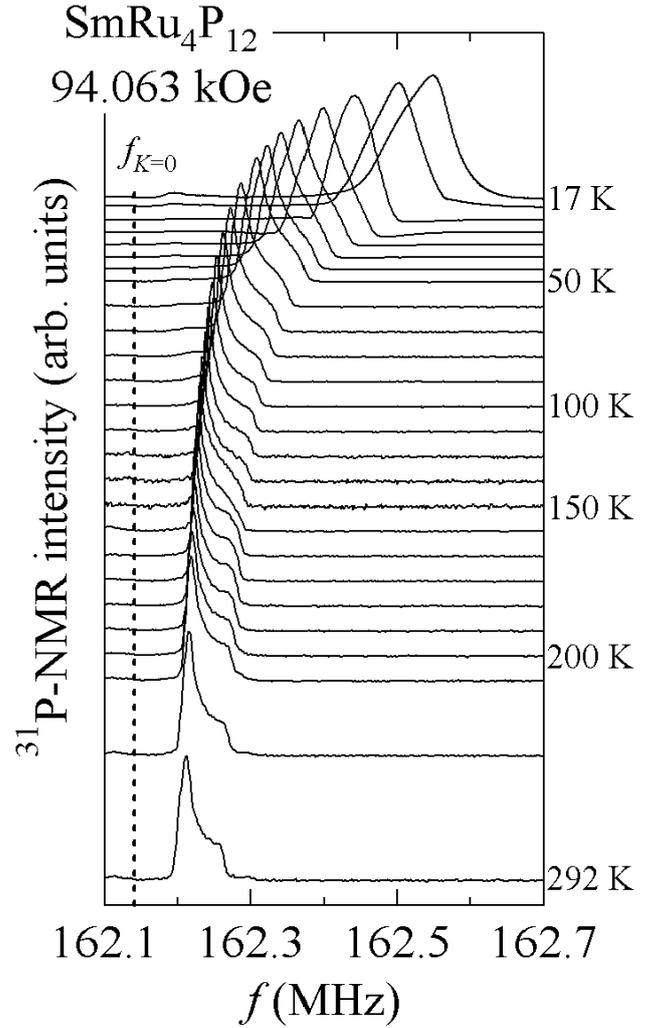

Fig. 1. Resonance frequency swept $^{31}$P-NMR spectra at several temperatures above $T_{MI}$ in the applied magnetic field of 94.063 kOe. The broken line shows the frequency of 162.14 MHz at $K = 0$ estimated from the nuclear gyro-magnetic ratio of the $^{31}$P nucleus.

### 3.2 $^{31}$P-NMR spectra below $T_{MI}$

The field swept $^{31}$P-NMR spectra at several temperatures below $T_{MI}$ at the frequencies of 9.23, 46.268 and 123.979 MHz are shown in Figs. 4 (a), (b) and (c), respectively. The shape of the spectra at 9.23 and 46.268 MHz (corresponding to the fields of about 5.5 and 27 kOe, respectively) have a triangle-like powder pattern. The previously obtained $^{31}$P-NMR spectra at 17 MHz also have a similar triangle-like feature.[17] The spectrum of SmRu$_4$P$_{12}$ is quite different from that of a simple AFM SmOs$_4$P$_{12}$ with $T_N$ of 4.6 K as seen in Fig. 4 (d). In the case of SmOs$_4$P$_{12}$ which has a homogeneous AFM structure, the spectra have a rectangle-type powder pattern.[20] The zero field $\mu$SR spectra of SmRu$_4$P$_{12}$ also suggest that the spin alignment is not so coherent.[14] The muon-spin precession was not clearly observed in SmRu$_4$P$_{12}$, whereas it was clearly observed in the case of SmFe$_4$P$_{12}$ and SmOs$_4$P$_{12}$ which have magnetically ordered ground states.[14,21] These results indicate the existence of an inhomogeneously distributed transferred hyperfine field in SmRu$_4$P$_{12}$, which reflects a non-simple magnetic structure such as an incommensurate structure.

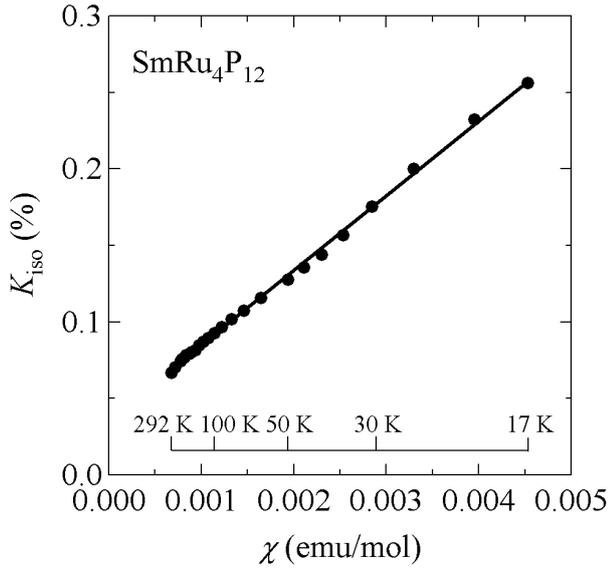

Fig. 2. Bulk susceptibility dependence of $K_{iso}$ ($K$-$\chi$ plot). The solid line shows the best-fit result by the formula (2). Some temperatures are also shown corresponding to $\chi$.

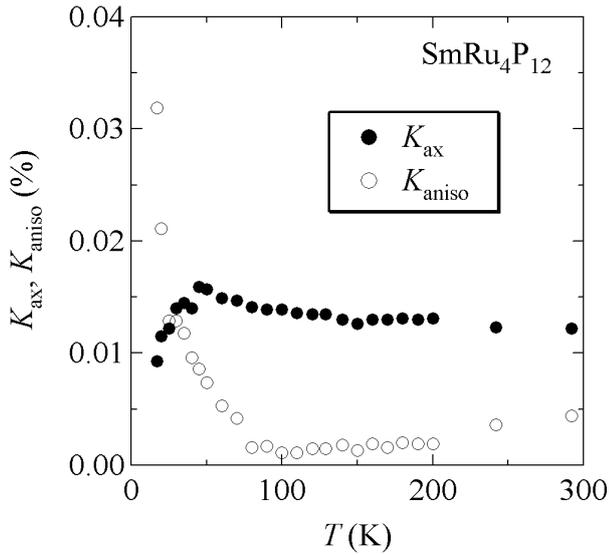

Fig. 3. Temperature dependence of $K_{ax}$ (closed circles) and $K_{aniso}$ (open circles).

Table I. Experimental and the dipolar calculated values of $K_{ax}$ and $K_{aniso}$ at the temperature of 292 K.

|  | $K_{ax}$ (%) | $K_{aniso}$ (%) |
| --- | --- | --- |
| Experimental | 0.012 | 0.0044 |
| Calculated | 0.007 | 0.0008 |

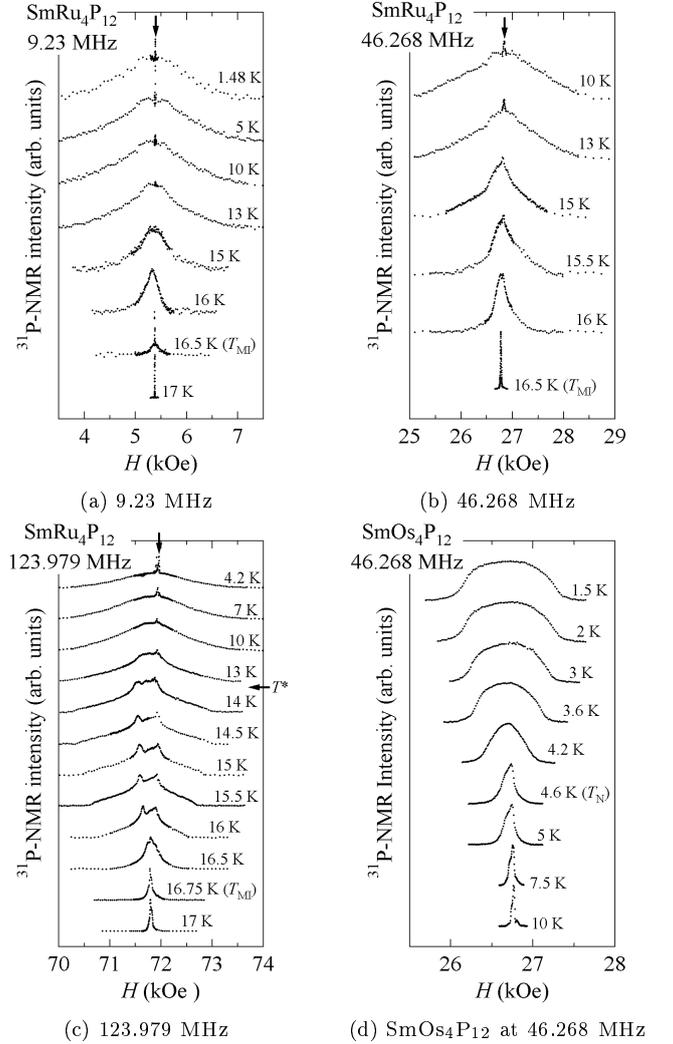

Fig. 4. Applied magnetic field swept $^{31}$P-NMR spectra of SmRu$_4$P$_{12}$ at several temperatures below $T_{MI}$ at the resonance frequencies of (a) 9.23, (b) 46.268 and (c) 123.979 MHz. The rather sharp signals pointed by the arrows in each figure arises from impurity phases with spurious $^{31}$P nuclei. The spectra of SmOs$_4$P$_{12}$ at 46.268 MHz are shown in (d).

With increasing field, the fine structure is induced by the fields around the center of the resonance line, as seen in the spectra at 123.979 MHz (corresponding to the field of about 72 KOe) in Fig. 4 (c). The shape of the spectra is not a simple triangle-like but much complicated pattern in the intermediate $T$ range between $T^*$ (about 13.5 K) and $T_{MI}$. Here, $T^*$ was determined from the anomalies

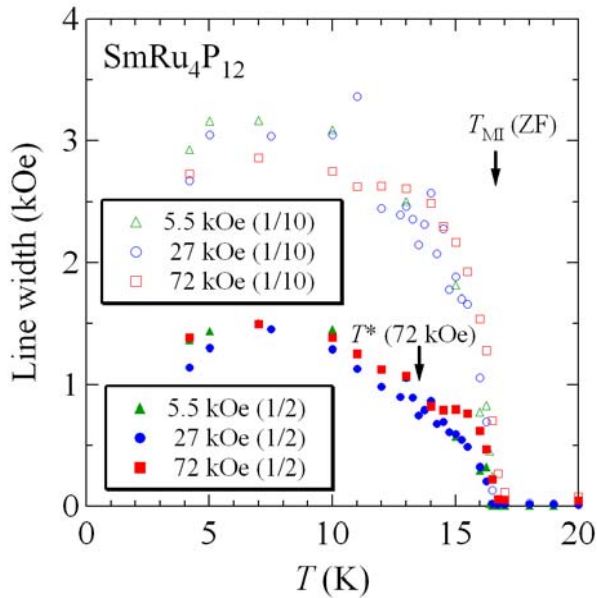

Fig. 5. Temperature dependence of the line width in several applied magnetic fields, which were estimated from the full width at half maximum (1/2) and at one-tenth maximum (1/10) values of the $^{31}$P-NMR spectra in Fig. 4 (a), (b) and (c), respectively.

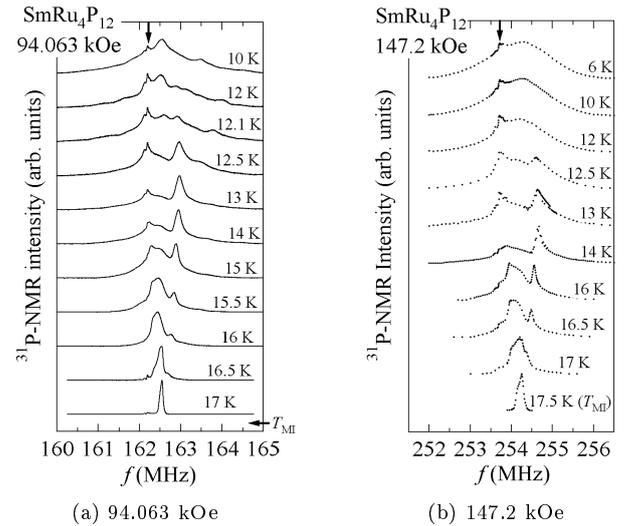

Fig. 6. Resonance frequency swept $^{31}$P-NMR spectra at several temperatures below $T_{\rm MI}$ in the applied magnetic fields of (a) 94.063 and (b) 147.2 kOe. The rather sharp signals pointed by the arrows in each figure arises from impurity phases with spurious $^{31}$P nuclei.

observed in $1/T_1$ which is discussed later in the next section.

Figure 5 shows the $T$ dependence of the line width in several fields estimated from the full width at half maximum and at one-tenth maximum values of the $^{31}$P-NMR spectra in Figs. 4 (a), (b) and (c), respectively. The line width rapidly increases below $T_{\rm MI}$, which indicates that an internal field at the $^{31}$P site develops just below $T_{\rm MI}$. The internal fields at 5.5 and 27 kOe are independent of the field in low fields. The $T$ dependence of the internal field at 72 kOe is different from that in low fields, which associates the appearance of the fine structure. No anomaly was observed in low fields around $T^*$. Since the field dependence is very small, the $T$ dependence of the internal field estimated from $^{31}$P-NMR is consistent with the previous $\mu$SR result which indicates the magnetic order below $T_{\rm MI}$.[14]

The field induced anomaly of the $^{31}$P-NMR spectra is enhanced in higher fields, as seen in Fig. 6, where the frequency swept spectra at the fields of 94.063 and 147.2 kOe (corresponding to the frequencies of about 162 and 254 MHz, respectively) $T_{\rm MI}$ are shown. In frequency swept spectra of wide resonance lines, the NMR signal often distorted owing to the technical reason, i.e., the small impedance mismatch of the network makes $rf$ reflection, which affects the shape of the spectra. In our case, the line width of 3-5 MHz is not so broad for getting the structure of the resonance lines. The separation of the fine structure below increases with increasing field.

A previously reported $^{31}$P-NMR result showed typical spectra with two trapezoid in shape below $T_{\rm MI}$,[22] which is different from the triangle-like one in our case. Our sample is a polycrystal synthesized by the high temperature and high pressure method,[8] and a grain of the powdered samples would include a lot of domains. Hence, the sample orients randomly to the fields in our measurement. Indeed, our $^{31}$P-NMR spectra above $T_{\rm MI}$ show a typical uniaxial powder pattern as seen in Fig. 1. On the other hand, the sample pieces of the previous report consiste of many small single crystals, and they were moderately crushed.[22] Then, the previously reported spectra would be obtained in a partially oriented sample condition. We consider that the difference of them arises from the crystal orientation to applied fields.

3.3 Spin-lattice relaxation rate $1/T_1$

Figures 7 and 8 show the $T$ dependence of $1/T_1$ in several fields. The $T$ dependence of $1/T_1$ of the nonmagnetic LaRu$_4$P$_{12}$ is also shown in Fig. 7 as a reference.[17] The $1/T_1$ of SmRu$_4$P$_{12}$ is nearly $T$ independent above $T_{\rm MI}$, where $1/T_1$ is dominated by the fluctuations of Sm$^{3+}$ 4f-localized moments. The $1/T_1$ rapidly decreases below $T_{\rm MI}$, and becomes nearly proportional to $T$ below about 3.5 K. The measured $1/T_1$ is the sum of the transferred contribution at the $^{31}$P site from Sm$^{3+}$ 4f-localized and from conduction electrons. In the low $T$ region ($1/T_1 \propto T$), the latter has a dominant contribution. The value of $1/T_1 T$ below 3.5 K is about 15 % of that of LaRu$_4$P$_{12}$, which indicates decrease of the density of states of the conduction electrons at $^{31}$P site associated with the M-I transition. The $T$ dependence of the $1/T_1$ is monotonous at low fields where no indistinct anomaly was observed around $T^*$. After subtracting the onsite ($^{31}$P site) contribution, the $T$ dependence of $1/T_1$ at low temperatures is nearly exponential. On the basis of the Raman scattering by two magnons, the energy gap of about 50 K was estimated.[23] This value is consistent with the previous $^{31}$P-NMR result and half of that obtained from an optical spectroscopy measurement.[17,24]

With increase field, the strong suppression of $1/T_1$ appears in the $T$ region between $T^*$ and $T_{\rm MI}$, as shown in

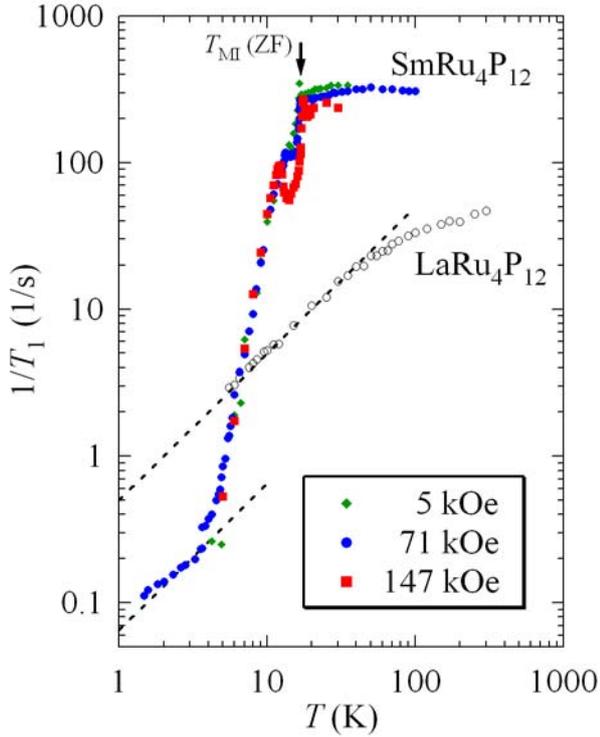

Fig. 7. Temperature dependence of $1/T_1$ in several applied magnetic fields. The $1/T_1$ of the non-magnetic LaRu$_4$P$_{12}$ is also shown as a reference.[17] The broken lines show guide to the eyes for the Korringa behavior ($1/T_1 \propto T$).

Fig. 8. The field induced phase transition at $T^*$ becomes apparent above 70 kOe,[22] and the anomaly enhanced with increasing field. At the field of 147 kOe, the critical slowing-down behavior were observed around both $T^*$ and $T_{\rm MI}$. Between $T^*$ and $T_{\rm MI}$ above 70 kOe where the spectra have the structure, $1/T_1$ were measured at the frequency of the line center of the spectrum just above $T_{\rm MI}$. The frequency locates between the two peaks of the structure, and corresponds to the peak of the triangle-like spectra below $T^*$. The behavior of $1/T_1$ is nearly the same as that obtained at two peaks of the structure.

## 4. Discussion

The $T$ dependence of both the $^{31}$P-NMR spectra and $1/T_1$ in various fields have demonstrated that the sharp anomaly at $T_{\rm MI}$ shifts to higher $T$ with increasing field. On the contrary, the round anomaly at $T^*$, which is undetectably small in low fields, becomes gradually apparent and shifts to lower $T$ with increasing field. The behavior is similar to that observed in other measurements.[9,10,12,13] The field-temperature ($H$-$T$) phase diagram obtained from $^{31}$P-NMR is shown in Fig. 9. The $T_{\rm MI}$ coincides with other data, whereas $T^*$ obtained from $1/T_1$ is lower than $T_{\rm N}$ corresponding to the anomalies obtained from other macroscopic measurements.[9,10] Though the reason of this behavior is not clear at the moment, the similar tendency also appears in the previous elastic constant result.[13] As a result, the $T^*$ obtained from $^{31}$P-NMR is considered to be $T_{\rm N}$.

The appearance of an internal field below $T_{\rm MI}$ in low fields shows that the phase transition at $T_{\rm MI}$ is a spontaneous magnetic transition but not an induced one. It is noted that the internal field obtained from $\mu$SR and from the low field $^{31}$P-NMR are essentially the same. A field induced anomaly around $T_{\rm N}$ were observed in both spectra and $1/T_1$ above 70 kOe. The $1/T_1$ between $T_{\rm N}$ and $T_{\rm MI}$ is sensitive to fields, whereas the effects of fields are small below $T_{\rm N}$.

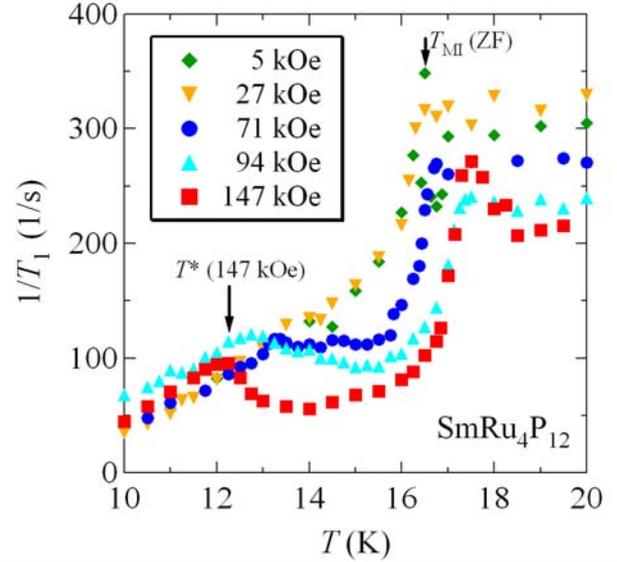

Fig. 8. Temperature dependence of $1/T_1$ in several applied magnetic fields around $T_{\rm MI}$ and $T_{\rm N}$, which was obtained by expanding Fig. 7.

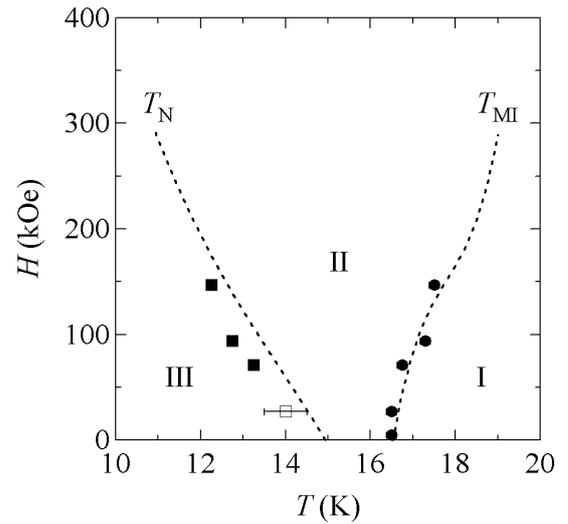

Fig. 9. $H$-$T$ phase diagram of SmRu$_4$P$_{12}$ obtained from $^{31}$P-NMR. The circles show $T_{\rm MI}$, and the squares show $T^*$. The open square means that the anomaly is not so apparent. The broken lines are obtained from macroscopic measurements.[10]

The magnetic transition occurs below $T_{\rm MI}$, where a dipolar or octupolar order is responsible for the phase transition. One of the aspects of this system is the complicated magnetic field response. An orbital contribution seems to have an important role. The ground state of SmRu$_4$P$_{12}$ is $\Gamma_{67}$ of the cubic $T_h$ symmetry,[19] which corresponds to $\Gamma_8$ state of the cubic $O_h$ symmetry. A scenario was proposed to account for the above phenomena with the second order transitions at $T_{\rm MI}$ and $T_{\rm N}$.[13] Among dipole, quadrupole and octupole moments in $T_h$ symmetry, the possible scenario is as follows. The phase III is a magnetic order of a dipole moment, which was well establish by $\mu$SR, $^{31}$P-NMR and a $^{149}$Sm nuclear resonant forward scattering measurement.[14,15,17,22,25] Then, a possible candidate for the order parameter of the phase II is the octupole moment of $T_i^\beta$ ($i = x$, $y$, $z$), which connects the phases II and III with the second order transition.[13] This model successfully explains the indistinct phase transition at $T_{\rm N}$ in low fields through the mixing of $\Gamma_{4u}$ and $\Gamma_{5u}$ in $T_h$ symmetry, which does not mix in the $O_h$ symmetry.

Further measurements including the neutron diffraction are highly expected for the further understanding of the complicated properties of SmRu$_4$P$_{12}$. The $^{31}$P-NMR with one single crystal is also expected to provide the accurate information for the complicated structure of the spectra.

## 5. Summary

In summary, $^{31}$P-NMR of the filled skutterudite SmRu$_4$P$_{12}$ has been carried out in several fields. The line width of the $^{31}$P-NMR spectra rapidly increases below $T_{\rm MI}$. In low fields, the effect of the field is small, and the internal field increases monotonously. On the contrary, the field induced anomaly becomes apparent with increasing field between $T_{\rm N}$ and $T_{\rm MI}$. The $T$ dependence of $1/T_1$ shows a gradual decrease in low fields, and successive anomaly at $T^*$ and $T_{\rm MI}$ in high fields.

**Acknowledgment**

The authors would like to thank T. Goto, M. Yoshizawa, K. Matsuhira, T. Hotta and S. Tsutsui for their useful discussions. This work was supported by a Grant-in-Aid for Scientific Research from the Ministry of Education, Sport, Science and Culture of Japan (no. 17740212). The work at Muroran Institute of Technology was supported by a Grant-in-Aid for Scientific Research in Priority Area "Skutterudite" (no. 15072201) of the Ministry of Education, Culture, Sports, Science and Technology, Japan.